# Strength and Sensitivity of Land-Atmosphere Interaction

Jun Yin[1]

[1]Department of Hydrometeorology, Nanjing University of Information Science and Technology

**Abstract**

The land-atmosphere coupling strength has been defined as the percentage of precipitation variability explained by the variation of soil moisture in the Global Land-Atmosphere Coupling Experiment (GLACE). While it is useful to identify global hotspots of land-atmosphere interaction, this coupling strength is different from coupling sensitivity, which directly quantifies how precipitation generation responds to the perturbation of soil moisture and is essential for our understanding of the global water cycle. To disentangle these two quantities, here we theoretically explore the relationships among coupling strength, sensitivity, and soil moisture variances. We use climate model outputs to show that the largest soil moisture variances are located in the transitional climate zones and the variations of soil moisture largely account for the geographical patterns of coupling hotspots. The coupling sensitivity is not necessarily low in non-hotspot regions, which could impose great impacts on the development of extreme climate events. We therefore call for more research attention on coupling sensitivity to improve our understanding of the climate system.

## 1. Introduction

Through exchanging energy, water, and momentum with the atmosphere, the land surface properties influence the weather patterns and global water cycle, playing a crucial role in the Earth's climate system. Various numerical experiments and observational studies have been conducted to understand this land-atmosphere coupling process, which is essential for improving weather prediction and climate projections. The pioneering work of the Global Land-Atmosphere Coupling Experiment (GLACE) has been designed to diagnose land-atmosphere coupling strength by comparing ensembles of coupled and uncoupled simulations, in which soil moisture is free to evolve or prescribed in each time step. It identifies typical land-atmosphere coupling hotspots of central Great Plains in North America, Sahel, and India, all of which are located in transitional climate zones.

When extended to the centennial timescale with slight variations in the experimental setup, the follow-up studies have shown that future transitional climate zones near central and eastern Europe also exhibit strong land-atmosphere coupling signals (Seneviratne et al. 2006). Since then, these experiments have been implemented in the climate models participating in the Coupled Model Intercomparison Project Phases 5 (CMIP5) models (Seneviratne et al. 2013) and further evolved into the Land Surface, Snow and Soil

Moisture Model Intercomparison Project (LS3MIP) endorsed by CMIP6, aiming to provide a comprehensive assessment of land-climate feedbacks in current climate models (Van Den Hurk et al. 2016).

Aside from GLACE experiments and their associated coupling metrics, various numerical and observational studies have been conducted at different timescales to investigate the land-atmosphere coupling. For example, the coupling strength has been quantified as the differences between the inter-annual variances of the mean seasonal temperatures in the coupled and uncoupled experiments (Lorenz et al. 2015); the coupling strength has also been quantified as the correlation coefficients between land properties and meteorological variables from observations and simulations (Seneviratne et al. 2006; Berg and Sheffield 2018); the coupling process has been split into two-stage interactions with intermediate variables of surface heat fluxes, each of the which has been quantified by the corresponding correlation coefficients (Dirmeyer et al. 2013). Other coupling metrics, such as mixing diagram and convective triggering potential, have been used to quantify the coupling process by using observational datasets (e.g., see a review by Santanello et al. 2018). While various metrics may have different measures of coupling strength, they tend to be consistent in the hotspots of the climate transitional regions.

Great efforts have been made to understand the links between the coupling strength and climate zones. It is argued that surface evaporation is nearly zero below the wilting point and reaches a plateau after the critical point, resulting in more sensitive responses of evaporation to the perturbation of soil moisture in transitional climate zones. This soil moisture-evaporation coupling pattern is further extended to the soil moisture-precipitation relationship. It is expected that precipitation is insensitive to soil moisture above a certain critical point over the wet regions where the evaporation reaches its maximum potential. Low evaporation in dry regions is expected to have limited impacts on precipitation. While this interpretation appears reasonable, it may confuse the concepts of sensitivity and strength of the land-atmosphere coupling. The former quantifies the response of atmospheric variables to the perturbation of soil moisture, whereas the latter refers to the variances of precipitation explained by variations of soil moisture. As explained later in this study, the geographical patterns of the coupling strength and sensitivity are not necessarily the same. The quantification of coupling sensitivity is as important as, if not more important than, the estimation of coupling strength for our understanding of the global water cycle and modeling of climate systems.

Toward this goal, here we start from the coupling metrics and theoretically explore the links between the coupling strength and sensitivity. We use cloud model outputs to find that soil moisture variance, being large in transitional climate zones, possibly accounts for geographical patterns of coupling strength. The estimated coupling sensitivity from numerical simulation may have large uncertainties in non-hotspot regions due to the limited ensemble size and smaller soil moisture variance. Such exploration underscores the importance of coupling sensitivity in controlling the global water cycle and regional water resources, which often has been overlooked in the studies of land-atmosphere coupling.

## 2. Large Soil Moisture Variations Amplify the Coupling Strength Signal

Identifying the contributions from local soil moisture and external forcing to precipitation generation is one of the challenges in diagnosing land-atmosphere coupling. In GLACE, this is done by averaging the precipitation over a relatively large size of ensemble simulations with prescribed soil moisture. In theory, the ensemble averages can smooth out the impacts of the external forcing if the ensemble size is infinite

$$\hat{P}_u = \lim_{n \to \infty} \frac{1}{n} \sum_{i=1}^{n} P_{u,i}, \tag{1}$$

which solely represents the contribution of prescribed soil moisture to the precipitation in the uncoupled experiments, $\hat{P}_u(s)$. Figure 1 shows how $P_{u,i}$ gradually converges to $\hat{P}_u$ with increasing ensemble size using a conventional stochastic weather generator (see Methods).

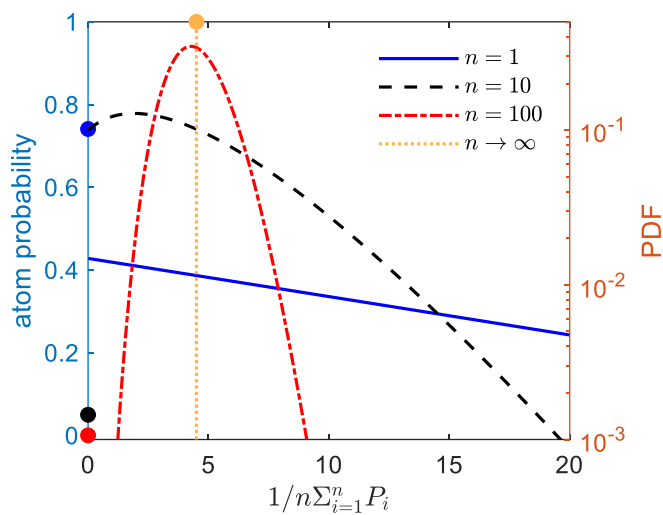

Fig. 1 Distribution of ensemble average of daily precipitation for prescribed soil moisture in the uncoupled experiments, $P_{u,i}$, with increasing ensemble size, $n$. The daily precipitation is a mixing of continuous and discrete distribution with an atom probability at zero, presenting days without rainfall.

In GLACE, the simulations are confined to the boreal summer (June, July, and August) to focus on the land-atmosphere coupling in the strongest season in the northern hemisphere. Assuming limited impacts of climate seasonality in these 3 months of

simulations and infinite ensemble size, one can determine the coupling strength, $\Delta\Omega$ (see Methods)

$$\Delta\Omega = \Omega_u = \frac{\sigma_{\hat{P}_u}^2}{\sigma_{P_u}^2} = \frac{\sigma_{\hat{P}_u}^2}{E[\sigma_{P(s)}^2]+\sigma_{\hat{P}_u}^2} \approx \frac{\left[\frac{d\hat{P}_u}{ds}\right]^2 \sigma_s^2}{E[\sigma_{P(s)}^2]+\left[\frac{d\hat{P}_u}{ds}\right]^2 \sigma_s^2}, \quad (2)$$

where $d\hat{P}/ds$ qualifies how $\hat{P}$ responses to soil moisture and should be interpreted as soil moisture – rainfall coupling sensitivity. This expression has the multiplication of the coupling sensitivity and soil moisture variance, suggesting both factors can play the same role in modulating the coupling strength. If the geographical variations of the coupling sensitivity are not strong, one may expect that the soil moisture variance may adequately account for the geographical patterns of coupling strength. This seems possible as the large soil moisture variations are often located in the transitional climate zones.

To address this point, we examine the global distributions of soil moisture variances in boreal summer from historical simulations of climate models (see Fig. 2a). As can be seen, the Great Plains of the United States, Sahel, and India, identified as GLACE hotspots, are also high in soil moisture variations. High values are also found in northern Asia, where the low radiation may suppress the moist convection and thus result in low coupling strength. Consistent with other independent studies, soil moisture variances typically increase before decreasing as regional aridity rises, leading to the maximum values in climate transition zones (see Fig. 2b). For GLACE, the soil moisture differences between the coupled and uncoupled experiments are large when the soil moisture variations are large. This is also evident in the latest LFMIP, where two typical examples are explicitly given in Fig. 2c and d. The historical simulations of soil moisture from the EC-Earth climate model have large seasonal and inter-annual variabilities over the hotspot of the southern Great Plains, displaying marked contrast to uncoupled experiments with climatological soil moisture. The soil moisture in a non-hotspot in Jiangxi China does not have strong inter-annual variabilities, resulting in synchronized changes in the coupled and uncoupled experiments. These soil moisture differences seem to be critical for interpreting the global land-atmosphere hotspots.

While the importance of soil moisture variability is stressed using the climate model outputs in Fig. 2, it is still difficult to disentangle the contributions from either soil moisture variability or coupling sensitivity. The latter, as addressed in detail in the next section, is usually not available in GLACE experiments and its estimation may suffer from background noise associated with external climate variability.

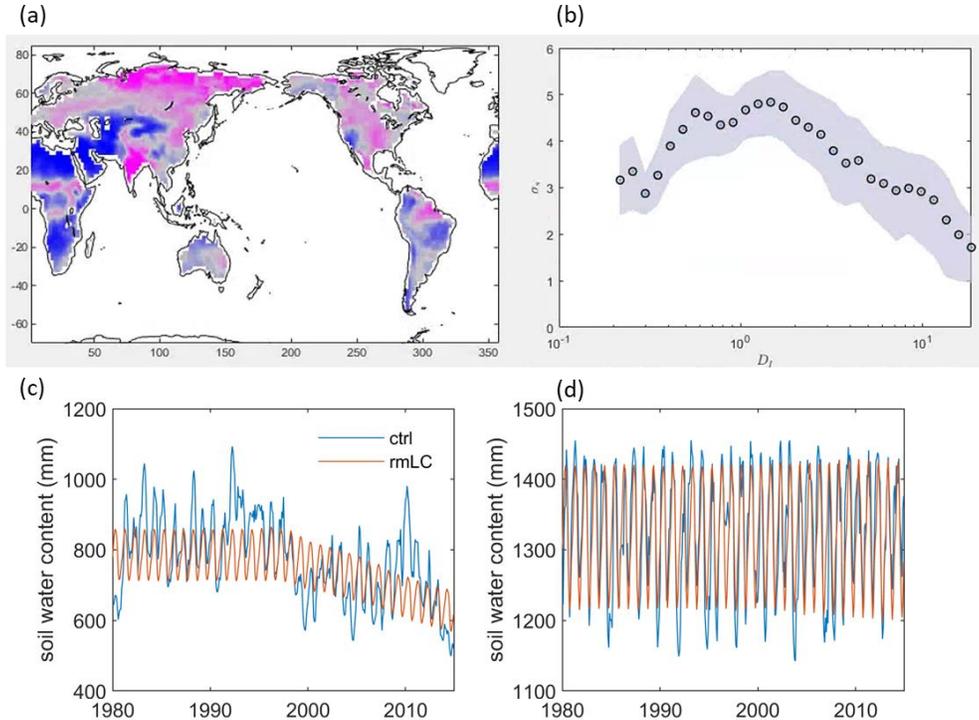

Fig. 2 (a) Global distribution of top-layer soil moisture variances from climate model outputs. (b) Relationships between soil moisture variances and local aridity; the shaded area represents the first and third quartiles. (c-d) Time series of monthly mean soil water content over the whole soil layers during 1980-2014 from EC-Earth climate models; the blue lines are from historical experiment and the red lines are from lfmip-rmLC experiment in the Southern Great Plain, USA and Jiangxi, China.

## 3. Large Background Noise Obscures the Coupling Sensitivity Signal

The first step to isolate the contributions to coupling strength is to quantify the soil moisture variability and coupling sensitivity. While the former is straightforward and has been presented in Fig. 2, the latter is challenging as there are no 'true' values to verify the estimation of coupling sensitivity. As pointed out by Koster et al., land-atmosphere coupling involves complex interactions between numerous processes and cannot be prescribed or parameterized in climate models (Koster et al. 2006). In this regard, we used a stochastic water balance model with a soil moisture–rainfall feedback module. This model integrates a root-zone water balance equation to a weather generator, and the feedback module allows us to directly parameterize the ensemble average precipitation, $\hat{P}(s)$ (see Methods). With prescribed coupling sensitivity, we inverse the problem of coupling sensitivity verification by exploring the statistical relationships between antecedent soil moisture and the following ensemble average daily precipitation.

In this stochastic model, ensemble average precipitation is the multiplication of rainfall intensity and frequency, where the latter is further parameterized as a linear function of soil moisture (see Methods). While the more sophisticated parameterization of soil moisture–rainfall feedback is available, the simplified version is used in this thought experiment with the hypothetical coupling sensitivity as we aim to qualitatively understand the typical patterns of land-atmosphere coupling in different climate zones.

With a moderate coupling sensitivity $d\hat{P}/ds = 3$ mm/day in all three typical climate zones, we can reproduce the patterns of soil moisture variability, which is low in wet or dry regions and large in climate transition zones (see Fig. 3). Due to the soil moisture boundaries at wilting point ($s = 0$) and near the field capacity ($s = 1$), soil moisture distributions tend to be positively/negatively skewed in dry/wet regions and become more symmetric in the transition zone, resulting in the nonlinear relationships between the soil moisture variances and local aridity (also see climate model outputs in Fig. 2).

Interestingly, by exploring antecedent soil moisture and following daily precipitation theoretically averaged over an ensemble of infinite size (see Methods), we found that the slopes between these two quantities are identical in all three climate zones, consistent with the prescribed coupling sensitivity of 3 mm/day (see circles and blue lines in Fig. 3). However, it is impossible to have infinite ensemble size from climate model simulations or observations. To mimic this limitation, we confine the total ensemble size to 200 in all three climate zones and theoretically quantify the first and third quantiles of ensemble average precipitation (see error bars in Fig. 3). The antecedent soil moisture is grouped into 12 bins of equal width, and the ensemble size in each bin is the multiplication of the total size and the probability of soil moisture within the bin. Therefore, one would expect larger sizes in the low end of the bins in dry regions as the soil moisture tends to be concentrated at the low level, whereas the size is small in the high end of the bins (sizes less than 5 are excluded in Fig. 3). The ensemble size controls the inter-quantile range (also see Fig. 1) and uneven sizes across different levels of soil moisture in dry or wet regions tend to have large variations in the quantile regression slopes (see shaded area in Fig. 3), suggesting large uncertainties in estimating coupling sensitivity in these non-hotspot regions.

From this thought experiment, we found that it is difficult to estimate the coupling sensitivities in wet or dry regions, which though could still be as large as those in coupling hotspots of the climate transition zones. The challenges lie in the background noise associated with the large chaotic forcing and small soil moisture variability, which overshadow the effects of land-atmosphere coupling. Statistically, the signal-to-noise ratios and coefficient of determination used to quantify the relationships between the antecedent soil moisture and following daily precipitation are low in wet and dry regions, further stressing the difficulty in quantifying the coupling sensitivities in non-hotspot regions. This offers a theoretical explanation for other independent studies which generally show consistent coupling strength from various metrics in climate transition zones but substantial disagreement in non-hotspot regions (Lorenz et al. 2015).

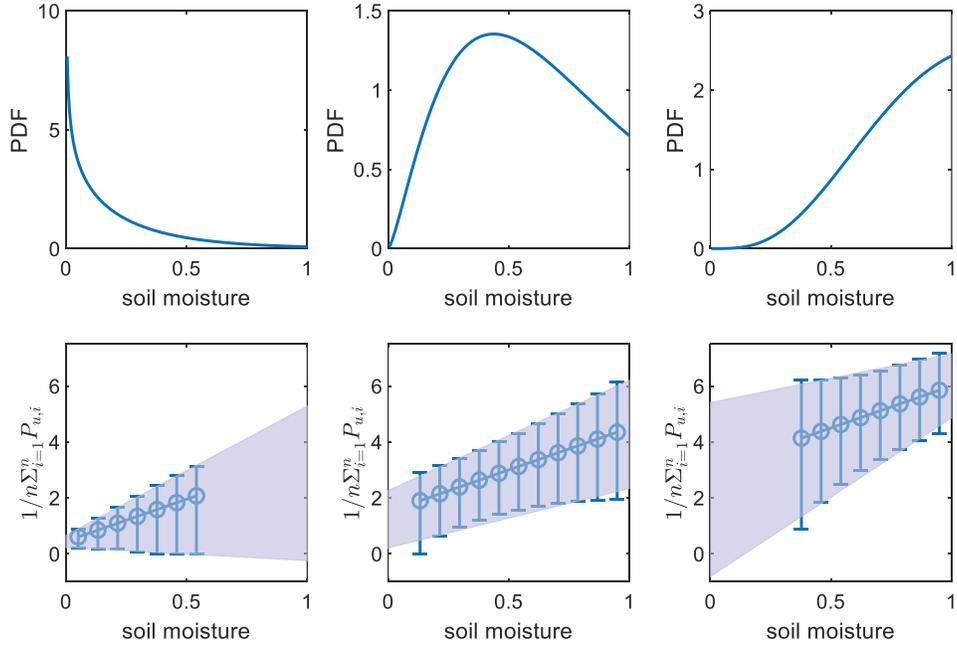

Fig. 3 (a-c) Relative soil moisture distributions over the typical (a) dry, (b) transitional, and (c) wet climate zones. (d-f) Relationship between prescribed soil moisture and daily rainfall rate under the same coupling sensitivity, $d\hat{P}/ds = 3$ mm/day. Rainfall is modeled by a stochastic weather generation with a soil moisture-rainfall feedback module (Yin et al. 2014).

## 4. Implications for Extreme Events

One may wonder whether it is still necessary to estimate the coupling sensitivity in non-hotspot regions. While this sensitivity may be less important under regular climate conditions when the coupling strength is low and the coupling signal is often obscured by large background noise, it may become particularly important when surface hydrological processes or large-scale climate conditions undergo significant changes. For example, drylands are usually non-hotspot regions but may have strong coupling sensitivity. If large-scale irrigation is applied in these drylands, the change in soil moisture, through the high coupling sensitivity, could influence regional precipitation. Similarly, soil moisture with low variability may experience significant changes due to land use land cover change or climate change, thus becoming important in modulating precipitation. In these cases, the coupling sensitivity, rather than the coupling strength, is the key to our understanding of the coupled hydroclimatic system.

To address this point, we again used the stochastic water balance model with a soil moisture-rainfall feedback module. We considered two different experiments with and without soil moisture –rainfall feedback. The feedback experiment allows the average rainfall frequency modeled as a function of soil moisture; the non-feedback experiment assumes the rainfall is completely controlled by the external forcing and thus the average rainfall frequency is set as a constant. It should be noted that this non-feedback experiment is different from the uncoupled GLACE experiment, which blocks the rainfall impacts on the soil moisture but still allows the influence of prescribed soil moisture on the boundary-layer dynamics and atmospheric convection.

Figure 4 compares the soil moisture dynamics under the feedback and non-feedback experiments in a typical wet region. The soil moisture distributions with and without feedback are very similar, showing negative skewness with modes at the saturation point (see Fig. 4a). This similarity suggests the land-atmosphere interaction has limited impacts on the overall soil moisture dynamics under regular climate conditions. However, when presented in the logarithmic scales (see Fig. 4b), the differences become clear and the probabilities of low soil moisture are much larger in the feedback model. The inter-arrival time of rainfall, being exponentially distributed under the non-feedback condition, has a heavier tail under the feedback conditions (see Fig. 4c), suggesting possible impacts of soil moisture – rainfall feedback on the dry-spell duration. The long drought duration could also influence the ecosystem's recovery from extreme drought events. Statistically, it takes a much longer time for soil moisture to increase from a low level to above the average value (see Fig. 4d). Moreover, the correlation coefficients between the soil moisture during a 15-day interval are larger under the feedback condition (see Fig. 4 e-f). Based on these analyses, it can be anticipated that soil moisture–rainfall feedback leads to longer and more persistent droughts, possibly impacting the ecosystem functionality and local water supply.

Given the significant impacts of land-atmosphere coupling on extreme hydro-climatic events, it is necessary to quantify the coupling sensitivity even in the non-hotspot regions where the coupling strength may be low. However, little attention has been paid to differentiating the quantities of coupling strength and sensitivity, which have been used interchangeably in the literature. Consequently, we may underestimate the impacts of land-atmosphere coupling in non-hotspot regions, where the coupling strength is low. For example, southeast China is not a hotspot as identified by GLACE; however, it may still have strong coupling sensitivity, which is significant for predicting persistent drought events, timing of the drought recovery, and ecosystem resilience.

Looking forward, we may need a new experiment setup different from GLACE to quantify the coupling sensitivity. This new experiment needs to prescribe soil moisture at different levels, preferably with some physical meaning such as wilting points, onset of water stress, field capacity, and saturation. With these experiments, one can estimate the coupling sensitivity, which provides direct information on how precipitation responds to the perturbation of soil moisture. For example, high coupling sensitivity suggests a great contribution to the rainfall deficit during drought events in wet climate zones, which though are identified as non-hotspots according to the low coupling strength. We therefore call for more research attention on the coupling sensitivity for the studies of land-atmosphere interaction.

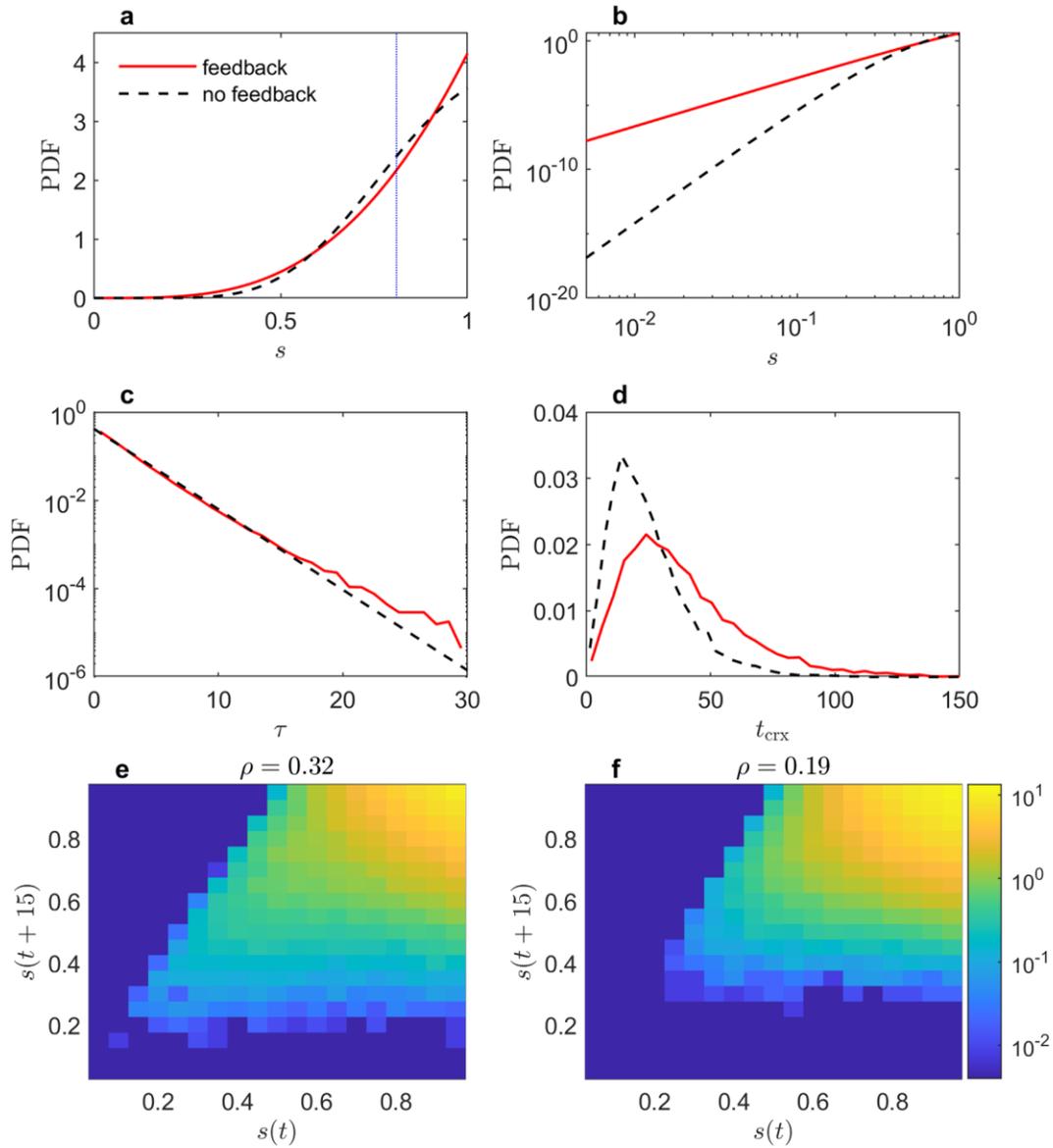

Fig. 4 Soil moisture dynamics with and without consideration of soil moisture – rainfall feedback from stochastic water balance model. (a, b) PDF of soil moisture in the regular and logarithmic scales; (c) PDF of inter-arrival time of rainfall; (d) PDF of soil moisture recovery time from 0.1 to the average value 0.81.

## Methods

### Rainfall rate conditional on prescribed soil moisture

The time series of rainfall events is conventionally assumed as a marked Poisson process of with a given average rainfall frequency, $\lambda$, and mean rainfall amount in each event, $\alpha$. For the prescribed soil moisture experiment, we assume the average rainfall frequency and rate are constant. The mean cumulated precipitation, $P$, over a period of $t$ is

$$\mu_P(\alpha,\lambda,t) = \alpha\lambda t, \tag{3}$$

the variance is

$$\sigma_P^2(\alpha,\lambda,t) = 2t\alpha^2\lambda, \tag{4}$$

and its distribution can be expressed as

$$f_P(P;\alpha,\lambda,t) = e^{-\lambda t}\left[e^{-\frac{P}{\alpha}}\sqrt{\frac{\lambda t}{x\alpha}}I_1\left(2\sqrt{\frac{\lambda Pt}{\alpha}}\right) + \delta(P)\right], \tag{5}$$

where $I_\nu(\cdot)$ is the modified Bessel function of the first kind of order $\nu$, and $\delta(\cdot)$ is the Dirac delta function.

### Stochastic water balance model

The minimalist soil water balance model with soil moisture rainfall frequency feedback

$$\lambda = as + b, \tag{6}$$

The pdf of soil moisture is

$$f_s(s) = \frac{(\gamma - a/\eta)^{b/\eta}}{\Gamma(b/\eta) - \Gamma(b/\eta, \gamma - a/\eta)}e^{(a/\eta - \gamma)s}s^{b/\eta - 1}, \tag{7}$$

the mean is

$$\mu_s = \frac{\eta[\Gamma(b/\eta + 1, \gamma - a/\eta) - \Gamma(b/\eta + 1)]}{(a - \gamma\eta)[\Gamma(b/\eta) - \Gamma(b/\eta, \gamma - a/\eta)]}, \tag{8}$$

and the variance is

$$\sigma_s^2 = \int_0^1 f(s)(s-\mu_s)^2 ds, \tag{9}$$

where can also be obtained analytically although the expression contains multiple gamma functions.

**The coupling strength**

The ensemble average of precipitation in the uncoupled experiments over an infinite ensemble size is essentially only controlled by the prescribed soil moisture, $\hat{P}_u(s)$. If the variance of soil moisture is known, the variance of its function can be derived from its Taylor expansion as

$$\sigma_{\hat{P}_u}^2 = \left[\frac{d\hat{P}}{ds}\right]^2 \sigma_s^2 - \frac{1}{4}\left[\frac{d^2\hat{P}}{ds^2}\right]^2 \sigma_s^4 + ..., \tag{10}$$

where $d\hat{P}/ds$ qualifies how $\hat{P}$ responses to soil moisture and should be interpreted as soil moisture – rainfall coupling sensitivity.

The Koster's coupling parameter is defined as

$$\Omega = \frac{n\sigma_{\hat{P}}^2 - \sigma_P^2}{(n-1)\sigma_P^2}, \tag{11}$$

where $n$ is the ensemble size. For $n \to \infty$, it becomes

$$\Omega = \frac{\sigma_{\hat{P}}^2}{\sigma_P^2}, \tag{12}$$

The coupling strength is defined as the differences of coupling parameters between coupled and uncoupled experiments

$$\Delta\Omega = \Omega_u - \Omega_c. \tag{13}$$

We first assume there is no seasonality, where the stochastic characteristics of large-scale forcing and land-atmosphere coupling strength do not change in the period of consideration. In the coupled experiments with infinite ensemble size, the ensemble average of precipitation at any time is the climatological mean value after the initial spin-up. If there is no seasonality, $\hat{P}_c(t)$ is constant in time and its variance is zero, leading to a zero coupling parameter in the coupled experiment, $\Omega_c = 0$.

In the uncoupled experiments with infinite ensemble size, $\hat{P}_u(s,t)$ at any time is only a function of the prescribed soil moisture, $\hat{P}_u(s)$, and its variance is linked to soil moisture variance as explained in Eq. (10). The distribution of precipitation rate with fixed soil moisture is $f_P(P;\alpha,\lambda)$ as given in Eq. (5). In GLACE experiment, the soil moisture is prescribed as one realization of the coupled process, which is expected to reach steady state after the initial spin-up and has the typical steady-state distribution $f_s(s)$. Under the land-atmosphere interaction, the ensemble average rainfall rate and depth are controlled by soil moisture, i.e., $\alpha(s)$ and $\lambda(s)$. The overall rainfall rate across all ensemble members and during an extended period can be described as a compound distribution $f_P(P;s)$, where the prescribed soil moisture $s$ is independent of external forcing for the precipitation generation and can be regarded as a parameter random variable. The variance of this compound distribution can be expressed as

$$\sigma^2_{P_u} = E[\sigma^2_{P(s)}] + \sigma^2_{\hat{P}_u}, \tag{14}$$

Substituting Eqs. (10) and (14) into (12) yields the coupling strength

$$\Delta\Omega = \Omega_u = \frac{\sigma^2_{\hat{P}_u}}{\sigma^2_{P_u}} = \frac{\sigma^2_{\hat{P}_u}}{E[\sigma^2_{P(s)}] + \sigma^2_{\hat{P}_u}} \approx \frac{\left[\dfrac{d\hat{P}}{ds}\right]^2 \sigma^2_s}{E[\sigma^2_{P(s)}] + \left[\dfrac{d\hat{P}}{ds}\right]^2 \sigma^2_s}. \tag{15}$$

The seasonal variations may be small within the three-month simulations in the GLACE but tend to increase the variances of $\hat{P}_c(t)$ and $\hat{P}_u(t)$ in both coupled and uncoupled experiments, leading to larger values in both $\Omega_u$ and $\Omega_c$. The differences between these two, defined as the coupling strength, reduce the impacts of seasonality on the quantification of the land-atmosphere coupling process.

**Acknowledgment**

Some results were utilized in the proposal submitted to the NSFC in 2024. This research is supported by the Natural Science Foundation of Jiangsu Province (BK20221343).